\documentclass[11pt, oneside]{article} 
\usepackage{afterpage}
\usepackage{epsfig}
\usepackage[a4paper]{geometry}
\usepackage{graphicx}	
\newcommand\ignore[1]{}			

\usepackage{amssymb}
\usepackage{amsmath}
\usepackage{amssymb}
\usepackage{amsthm}
\usepackage{graphicx}
\usepackage[urlcolor=blue,colorlinks=true,linkcolor  = black]{hyperref}
\usepackage{placeins}
\usepackage{tocloft}
\usepackage{slashed}
\usepackage{float}
\usepackage{fullpage}
\usepackage{braket}
\usepackage{url}

\usepackage{tabularx}

\usepackage{cite}
\usepackage{subcaption}
\usepackage{upgreek}
\usepackage{outlines}
\usepackage{braket}
\usepackage{appendix}
\def\0{{(0)}}
\def\1{{(1)}}
\usepackage{xspace}

\usepackage{mathrsfs}

\usepackage{bm}

\usepackage{caption}
\usepackage{subcaption}

\def\ccccend{\end{array}\right)}

\usepackage{tikz}
\usepackage{pgfplots}
\usetikzlibrary{matrix,arrows,decorations.pathmorphing}
\usetikzlibrary{patterns}
\usetikzlibrary{shapes.misc}
\usetikzlibrary{trees}
\usetikzlibrary{decorations.pathmorphing}
\usetikzlibrary{shapes.geometric}
\usetikzlibrary{positioning}
\usetikzlibrary{decorations.markings}
\usetikzlibrary{positioning,arrows}

\usetikzlibrary{decorations.pathreplacing}
\usetikzlibrary{decorations.pathmorphing}
\usetikzlibrary{shapes}
\usetikzlibrary{arrows.meta}
\usetikzlibrary{plotmarks}
\usetikzlibrary{decorations.markings}
\tikzset{
particle/.style={thin,draw=black, postaction={decorate},
decoration={markings,mark=at position .5 with {\arrow[black, line width=0.5mm]{stealth}}}},
gluon/.style={decorate, draw=black, decoration={coil,amplitude=4pt, segment length=5pt}},
photon/.style={decorate, decoration={snake}},
singularity/.style={decorate, draw=black, decoration=zigzag}
}

\usepackage[bottom]{footmisc}

\theoremstyle{definition}

\usepackage{amsthm}
\theoremstyle{theorem}

\begin{document}


\makeatletter
\@addtoreset{equation}{section}
\makeatother
\renewcommand{\theequation}{\thesection.\arabic{equation}}

\begin{center}
\phantom{a}\\
\vspace{0.8cm}
\scalebox{0.90}[0.90]{{\fontsize{24}{30} \bf{Axions in the Dark Dimension}} }\\ 
\end{center}

\vspace{0.4cm}
\begin{center}
\scalebox{0.95}[0.95]{{\fontsize{15}{30}\selectfont Naomi Gendler and Cumrun Vafa}}
\end{center}

\begin{center}
\vspace{0.25 cm}
\textsl{Jefferson Physical Laboratory, Harvard University, Cambridge, MA 02138 USA}\\

	 \vspace{1.1cm}
	\normalsize{\bf Abstract} \\[8mm]
\end{center}
\begin{center}
	\begin{minipage}[h]{15.0cm}
The dark dimension scenario, which is motivated from Swampland principles and predicts a single micron scale extra dimension, suggests a consistent framework for the dark sector of the universe.
We consider the implications of this scenario for the QCD axion.  We find that in the scenario in which the axion is localized on the standard model brane (which we will argue is natural), a combination of theoretical (being bounded by the 5D Planck mass) and observational constraints forces it to have decay constant in a narrow range $f\sim 10^9-10^{10}{\, \rm  GeV}$. This corresponds to a mass for the QCD axion of $m_a\sim (1-10)\ {\rm meV}$.  The axion mass surprisingly coincides with the mass scale for the dark energy, the dark matter tower,  and the neutrinos.  In this scenario axions are not expected to form a large fraction of the dark matter but nevertheless this range of axion parameters is accessible to observations in near future experiments.   

	\end{minipage}
\end{center}
\vfill
\today
\newpage

\section{Introduction}
The dark dimension scenario has been recently proposed as a way to reconcile the smallness of dark energy with Swampland principles \cite{Montero:2022prj}.  It leads to the prediction of one extra mesoscopic dimension in the range of 1-10 microns.  This also suggests a unified framework connecting dark energy to dark matter \cite{Gonzalo:2022jac} with interesting phenomenology \cite{Law-Smith:2023czn,Obied:2023clp} (see \cite{Anchordoqui:2024akj} for other approaches to dark dimension phenomenology).  For a review see \cite{Vafa:2024fpx}.

One of the hopes of the Swampland program is to alleviate the fine tunings in the phenomenological models of our universe, by appealing to consistency principles dictated by quantum aspects of gravity.
Some of the glaring fine tunings that we observe in the universe are: the smallness of  dark energy, the smallness of the electroweak scale, and the smallness of the QCD $\theta$-angle, also known as the strong CP problem.
In the dark dimension scenario, while the smallness of dark energy is not explained, combining the observation of its smallness with the expectations of various Swampland principles leads to the prediction of an extra dimension of size $R \sim 1-10$ microns. In this scenario, this scale is related to dark energy by $R\sim \Lambda^{-1/4}$.  Furthermore, a heuristic argument \cite{Montero:2022prj,Gonzalo:2022jac} links the standard model scale with  $\Lambda^{1/6}$ (where we have set the Planck mass to 1).  The aim of this paper is to investigate the observational consequences of a QCD axion (i.e. an axion solving the strong CP problem) in this scenario.

A natural solution of the strong CP problem is to introduce an axion via the  Peccei-Quinn mechanism \cite{Peccei:1977hh}.  That axions should exist is clear from the Swampland perspective: there are no free parameters in quantum gravity, and therefore all parameters are expectation values of dynamical fields.  We can then ask what the couplings of these dynamical fields are to the standard model. In particular, the coupling of the QCD axion to photons is governed by the decay constant parameterizing its kinetic term:  $ {1\over 2}\int f_a^2|\nabla \theta|^2$.  The mass of the QCD axion is also fixed by the decay constant: $m_a\sim \Lambda_{QCD}^2/f_a$, where $\Lambda_{\text{QCD}} \approx \sqrt{m_\pi f_\pi} $ is the scale of the QCD confinement effects that generate the potential for the axion.    In the dark dimension scenario, the Standard Model is localized on a brane that is codimension one in the five dimensional spacetime.  In such a scenario, as we will explain, it is natural to consider the QCD axion also being bound to this brane, though we will also consider the case in which the axion propagates in the bulk of the fifth dimension.  Our main insight comes by applying the Weak Gravity Conjecture (WGC) \cite{Arkani-Hamed:2006emk} to the axion, supported by string theory examples.  Namely, this conjecture states that
\begin{align}
f_a\lesssim M_p.
\label{eq:fboundintro}
\end{align}
In the general form of the WGC, the $M_p$ inequality above refers to the Planck scale in the uncompactified spacetime.  However, we will argue that when applied to axions localized on a brane, the relevant scale is the Planck scale in the bulk theory where the axion is localized.  Using this generalized notion, the above inequality refers to the five dimensional Planck scale, $M_5\sim 10^{9}-10^{10}{\, \rm GeV}$.  Experimental bounds on the coupling of the QCD axion to photons requires $f_a \gtrsim 10^{9}{\, \rm GeV}$ \cite{Lella:2023bfb, Chakraborty:2024tyx}. Combining this bound with \eqref{eq:fboundintro} leads to a prediction of
\begin{align}
    f_a\sim 10^9-10^{10}{\, \rm GeV}
    \label{eq:fpredintro}
    \end{align}
and an axion mass of
$$m_a\sim {\Lambda_{\rm QCD}^2\over f_a} \sim 1-10\ {\rm meV}.$$
This range of decay constants alleviates the problem of overclosing the universe, pointed out in \cite{Preskill:1982cy,Abbott:1982af,Dine:1982ah}. If $f_a\gtrsim 10^{11}-10^{12}{\, \rm GeV}$, assuming $\mathcal{O}(1)$ initial misalignment of the QCD axion and standard reheat temperatures, too much dark matter is produced.  Without the introduction of an intermediate scale between the TeV and the GUT or Planck scale (such as, in our case, the five-dimensional Planck scale),  arranging for a small enough decay constant would appear as a fine tuning problem. For simple string compactifications, it is indeed rather difficult to obtain such a low value for $f_a$, although see \cite{Demirtas:2021gsq, Gendler:2023kjt, Mehta:2021pwf, Demirtas:2018akl} for obtaining small enough values of $f_a$ in compactifications with large numbers of axions. The scenario described here realizes the mechanism in \cite{Svrcek:2006yi,Conlon:2006tq} of using a localized QCD axion to ensure a decay constant in the allowable range of parameters.

Another motivation for this work, independent of the Swampland reasoning for the dark dimension setup, is simply to ask what the constraints are on a QCD axion in a theory with a single large extra dimension on the boundary of what has been excluded with observation. Previous studies of axions in theories with extra dimensions primarily focus on setups aiming to solve the electroweak hierarchy problem \cite{Dienes:1999gw,DiLella:2000dn,Ma:2000rh,Horvat:2003xd,Collins:2002kp}. The Planck scale of the extra dimensional bulk was therefore much lower than the scale in the dark dimension scenario we consider here. As a consequence, these models were forced to have at least two large extra dimensions, while in the dark dimension scenario the number of large extra dimensions is exactly one. 

The assumption of previous works, that the higher-dimensional Planck mass was at the TeV scale, led them to assume that the axion decay constant was also TeV-scale. Having a decay constant this low then \textit{excludes} the possibility that the axion is localized on the Standard Model brane, since the four dimensional axion decay constant would then also be TeV-scale, which is severely ruled out by observations. In the present work, a localized axion is not ruled out, and furthermore is a natural option within reach of future axion-detection experiments.

The rest of this paper is organized as follows. In \S\ref{sec:review} we review some salient features of axions arising in string theory compactifications and relate them to the WGC. In \S\ref{sec:expbounds} we briefly review phenomenological constraints on axions.  In \S\ref{sec:axionDD} we discuss the constraints on axions in the dark dimension scenario, first in the case that it is localized on the Standard Model brane, and subsequently in the case that it is a bulk field.

\section{Review of the Relevant Axion Physics} \label{sec:review}
The QCD axion was originally proposed to solve the strong CP problem. The QCD Lagrangian is allowed to have a CP-violating term:
\begin{align}
    \mathcal{L} \supset \frac{ \theta}{32\pi^2} \epsilon^{\alpha \beta \mu \nu} \text{tr} G_{\alpha \beta} G_{\mu \nu}
\end{align}
where $\theta$ parameterizes the amount of CP-breaking. Experiments on the neutron electric dipole moment tell us that $\theta \lesssim 10^{-10}$ \cite{Abel:2020pzs,Baker:2006ts,Pendlebury:2015lrz}. The question of why this number is so small is known as the strong CP problem.

Several solutions to the strong CP problem have been proposed, but arguably the most attractive one is the idea of Peccei and Quinn \cite{Peccei:1977hh} to promote the parameter $\theta$ to a dynamical field---an axion. This axion has a potential that is generated by QCD effects and takes the form
\begin{align}
    V_{\text{axion}} \supset \frac{1}{2} \Lambda_{\text{QCD}}^4 \frac{m_u m_d}{(m_u+m_d)^2} \theta^2  + \mathcal{O}(\theta^4).
\end{align}
If this term is the only significant contribution to the QCD axion potential, then the axion will dynamically relax to $0$, providing a mechanism for the observed smallness of $\theta$.

As a dynamical field, the QCD axion also has a kinetic term
\begin{align}
    \mathcal{L} \supset \frac{1}{2} f_a^2 \partial_\mu \theta \partial^\mu \theta.
\end{align}
Canonically normalizing this field, then, the mass of the QCD axion, $a$, is:
\begin{align}
    m_a \approx \frac{\Lambda_{\text{QCD}}^2}{f_a}.
    \label{eq:qcdmass}
\end{align}

The QCD axion can couple to Standard Model fields in addition to gluons. In particular, it can couple to photons through the CP-violating term in the QED Lagrangian:
\begin{align}
    \mathcal{L} \supset \frac{g^2 a}{16\pi^2 f_a} F \wedge F
\end{align}
where  $F$ is the electromagnetic field strength. The size of $f_a$, then, sets the strength of the coupling between the QCD axion and photons. As we will discuss in \S\ref{sec:expbounds}, values of $f_a$ that are too small are excluded because of this coupling by bounds on astrophysical processes and direct detection experiments. Note that the coupling of the QCD axion to photons is not a further assumption about the axion model: rather, this coupling is induced via mixing with pions (see \cite{Hook:2018dlk} for a pedagogical review).

In string theory, axions are pervasive. One way to obtain an axion in string theory is to note that higher-dimensional gauge fields integrated over cycles in a compactification give rise to axion-like particles in lower dimensions. Consider, for example, string theory compactified on a six-dimensional manifold, $X$. In ten dimensions, string theory has various $p$-form gauge fields (which we will denote $A_p$ for generality), which when reduced on a $p$-cycle $\Sigma_p$ give rise to axions in the lower-dimensional theory. Concretely, we begin with a ten-dimensional action that includes a gauge kinetic term for $A_p$ (following the conventions of \cite{Svrcek:2006yi}):
\begin{align}
    S_{10D} \supset -\frac{1}{2} \frac{2\pi}{\ell_s^{8-p}} \int d^{10}x |dA_p|^2.
\end{align}
We can expand the gauge fields as
\begin{align}
    A_p = \frac{1}{2\pi} \sum_I \theta_I \omega_I
\end{align}
where
\begin{align}
    \int_{\Sigma_p^I} \omega_J = \delta_{IJ}
\end{align}
and the axions $\theta_I$ are defined as
\begin{align}
    \theta_I = \int_{\Sigma_p^I} A_p.
\end{align}
Dimensionally reducing down to four dimensions, we obtain the kinetic term for the axions
\begin{align}
  \mathcal{L}_{\text{kin}}= -\frac{1}{2} \left[ \frac{1}{\ell_s^{8-2p}} \int_X \, \omega_p^I \wedge \star \omega_p^J \right]  \partial_\mu \theta_I \partial^\mu \theta_J.
\end{align}
Using the fact that the the 4D Planck mass satisfies $M_p^2 = \frac{1}{\ell_s^8} \mathrm{vol}(X_6)$, and setting a length scale $r$ for $X_6$ such that $\mathrm{vol}(X_6) \sim r^6$, we have that the axion decay constants have the scaling\footnote{By setting such a length scale for $X_6$ we are implicitly assuming that $X_6$ is roughly isotropic. In explicit scans where $X_6$ is a (generally non-isotropic) Calabi-Yau threefold, experimentally one finds that the scaling of the decay constants for axions coming from the $C_4$ gauge field in type IIB is better approximated by $\frac{f}{M_p} \sim \mathrm{vol}(X_6)^{-1/2} \mathrm{vol}(\Sigma_4)^{-1/4}$, where volumes are measured in string units \cite{JunyiWIP}.} \cite{Svrcek:2006yi}
\begin{align}
    \frac{f_a}{M_p} \sim \frac{1}{(r/\ell_s)^p}.
    \label{eq:fscaling}
\end{align}

This simple example illustrates a general point: compactifications of higher-dimensional theories with gauge fields lead to axions in lower dimensions. String theory compactifications, then, generically give rise to axions. We also see here that the decay constant of the axion is determined by the internal geometry of the compactification.

In the spirit of the above example, if we were to naively integrate a higher form gauge field over a cycle of some internal 5-dimensional manifold to arrive at a theory realizing the dark dimension scenario, we would obtain an axion that propagates in the bulk of the large extra dimension. We can also, however, obtain axions from string theory that are localized in some direction. Such localized axions can be obtained if the axion comes from reducing a gauge field of sufficiently high form on a blow-up mode of a manifold.

We will also make use of the Weak Gravity Conjecture (WGC), and as such we will briefly review it here. The WGC states that in a theory with a U(1) gauge field, there should exist a particle whose charge-to-mass ratio is bigger than or equal to the charge-to-mass ratio of an extremal black hole in the theory:
\begin{align}
    \frac{Q}{m} \geq \left(\frac{Q}{m}\right) \bigg|_{\text{ext. BH}}
\end{align}

One might expect that a similar bound exists for theories with arbitrary $p-$form gauge fields. In particular, such a bound would relate 0-form gauge fields (i.e. axions) to instantons.  The axion form of the WGC reads \cite{Arkani-Hamed:2006emk}
\begin{align}
    f_d \lesssim  M_d,
\end{align}
where $f_d$ and $M_d$ are the axion decay constant and Planck mass in $d$ dimensions, respectively.  We will make use of this bound in \S\ref{sec:axionDD}.

\section{Phenomenological constraints on axions} \label{sec:expbounds}
There are many ongoing and planned experiments looking for signals of the QCD axion and axion-like particles. There are experiments that probe couplings of axions to various standard model particles (see e.g. \cite{Capozzi:2020cbu,Buschmann:2021juv, Esser:2023fdo}), but the observational constraints that we will be focused on are those having to do with couplings of axions to photons, as well as with axion misalignment dark matter. We will now review the current constraints and future experiments that probe these quantities. In \S\ref{sec:axionDD} we will compare estimates on axions in the dark dimension scenario to these constraints.

Experiments probing axion-photon couplings are sensitive to specific axion masses, usually spanning some range. Axion-photon coupling experiments fall into the categories of direct detection and indirect detection. 

Direct detection experiments rely on direct axion-photon conversion in the presence of a magnetic field, and hope to detect axions by converting a source of some axion flux to photons by setting up a large magnetic field. Examples of such setups are solar axion experiments such as CAST, and resonance-based detectors (which rely on an axion of a given mass making up an $\mathcal{O}(1)$ fraction of dark matter). The most broadband current bound is given by CAST \cite{CAST:2017uph},  sensitive to axion masses  $m_a \lesssim 0.02 \ \text{eV}$ and tells us that $g_{a\gamma \gamma} \lesssim 6.6 \times 10^{-11} \ \text{GeV}^{-1}$ (i.e. $f_a \gtrsim 1.76 \times 10^7 \ \text{GeV}$).

Indirect detection experiments focus on astrophysical sources and consider stellar evolution in the presence or absence of an axion. The most relevant general bound comes from looking at globular clusters, and noting that the presence of axions would affect the lifetime of stars within these clusters. These observations give a bound \cite{Ayala:2014pea,Dolan:2022kul} $g_{a\gamma \gamma} \lesssim 4.7 \times 10^{-11} \ \text{GeV}^{-1}$ (i.e. $f_a \gtrsim 2.47 \times 10^7 \ \text{GeV}$) for axions in the mass range $10^{-6}\lesssim m_a \lesssim 1 \ \text{eV}$. 

The above experiments constrain general axion-like particles, but in the case of the QCD axion, the bounds can be much stronger. Stringent bounds on the QCD axion decay constant come from requiring that neutron stars do not cool too quickly \cite{Buschmann:2021juv}. These constraints lead to the limit 
\begin{align}
f_a \gtrsim 3.5 \times 10^{8} \ \text{GeV} \ \  (\text{i.e. } g_{a\gamma \gamma} \lesssim 3.3 \times 10^{-12} \ \text{GeV}^{-1}).
\end{align}
Another very relevant recent bound comes from observing energy losses in supernovae \cite{Lella:2023bfb,Chakraborty:2024tyx}. They find that for the QCD axion $f_a \gtrsim 1.8 \times 10^9 \, \text{GeV} \ \ \text{(i.e. } g_{a\gamma \gamma} \lesssim 6.4 \times 10^{-13} \ \text{GeV}^{-1}).$
This bound already probes the parameter regime suggested by \eqref{eq:fpredintro}.

Finally, one can consider the constraint that more dark matter than we observe would overclose the universe. Axions can contribute to the overall dark matter density due to coherent axion states oscillating in their potentials in the early universe. The equation of motion for such a state is that of a damped harmonic oscillator, where the friction term is proportional to the Hubble constant, $H$, and the spring constant is proportional to the axion mass, $m_a$. When $H$ decreases below $m_a$, the system transitions from an overdamped to an underdamped oscillator, and the axion state begins to oscillate and contribute to the energy density of the universe. This contribution therefore depends on the initial misalignment angle at the time that $H \approx m_a$, the axion mass, and importantly the width of the potential, which is controlled by $f_a$. Assuming $\mathcal{O}(1)$ initial misalignment angle and a standard cosmological history, the bigger $f_a$ is, the more dark matter there will be. For the QCD axion, whose mass is given in \eqref{eq:qcdmass}, the dark matter density is given by \cite{Marsh:2015xka}
\begin{align}
    \Omega_a h^2 \sim 2 \times 10^4 \left(\frac{f_a}{10^{16} \ \text{GeV}} \right)^{7/6} \langle \theta_{\text{initial}}^2 \rangle
\end{align}
where $h$ is the dimensionless Hubble parameter. Assuming $\theta_{\text{initial}} \sim \mathcal{O}(1)$, this leads to the bound on $f_a$:
\begin{align}
    f_a \lesssim 10^{11}-10^{12} \ \text{GeV}.
\end{align}

The most promising upcoming experiment that will probe axion-photon couplings in the region we are interested in is the International Axion Observatory (IAXO) experiment  \cite{IAXO:2019mpb}. This experiment would look for axion-like particles coming from the sun, having been produced via the Primakoff process. The experiment does not assume anything about the proportion of dark matter made up by axions, only on the production mechanism that generates a flux of solar axions. IAXO projects to reach a sensitivity of $g_{a \gamma \gamma} \sim 10^{-12} \, \text{GeV}^{-1}$, or in other words a QCD axion mass of
\begin{align}
    m_a \sim 1 \, \text{meV}- 1 \, \text{eV},
\end{align}
putting bounds on the QCD axion decay constant of
\begin{align}
    f_a \gtrsim  10^{9} \, \text{GeV}.
\end{align}

It is worth noting that BabyIAXO \cite{IAXO:2020wwp} is set to begin operations in the next few years and will already probe axion-photon couplings an order of magnitude lower than CAST.

\section{Axions in the dark dimension scenario} \label{sec:axionDD}

Let us assume that there exists a single large extra dimension with a five-dimensional Planck mass $M_5 \approx 10^9- 10^{10}$ GeV. This is motivated by the dark dimension scenario \cite{Montero:2022prj}. In this setup, a large extra dimension is implied by the notion that the smallness of the cosmological constant is a consequence of realizing our universe as a string theory solution in an asymptotic region of field space. If this is the case, then there should exist a tower of states whose mass is set by the value of the cosmological constant:
\begin{align}
    m_{\text{tower}} \sim \Lambda^{\alpha}
\end{align}
with $\alpha$ an $\mathcal{O}(1)$ number taken to satisfy  $\frac{1}{4} \leq \alpha \leq \frac{1}{2}$ \cite{Montero:2022prj}. Bounds on deviations from Newton's law force $\alpha \approx \frac{1}{4}$ in this range. The only experimentally allowed model of large extra dimensions compatible with a tower of masses with characteristic scale $\Lambda^{1/4}$ is a single large extra dimension with length scale $R \approx 1-10  \, \mu\text{m}$, and corresponding Planck mass
\begin{align}
    M_5 \approx 10^{9} - 10^{10} \ \text{GeV}.
\end{align}
Our investigation therefore boils down to asking about constraints on a QCD axion in a setup with one large extra dimension with Planck mass given as above.

In this scenario, the Standard Model fields must be realized on a brane localized in 5D, otherwise we would get a tower of light particles for each Standard Model field.   We will not consider what effects might stabilize the extra dimension in this scenario, and leave to the future the work of analyzing whether such a setup can be realized in an explicit string theory construction in a way that accords with all observational bounds. 

In this work, we will simply ask the question of whether a QCD axion in such a scenario is allowed experimentally.  As such, we will assume that there exists a QCD axion in this scenario that does indeed solve the strong CP problem. We will consider both the case that the QCD axion is localized to the Standard Model brane, which we will argue is more natural, as well as the case that the axion propagates in the bulk of the fifth dimension. We will see that in the case that the QCD axion is localized on the brane, using an assumption about decay constants of axions in five dimensions leads to a prediction for the QCD axion decay constant that is within reach of future experiments.

\subsection{Case 1: Axions localized on 4D brane}

We first consider the case where the QCD axion is localized to the Standard Model brane in the large extra dimension. This could happen, for example, if the QCD axion is the complex part of a blow-up mode in the string theory context. It was argued in \cite{Heckman:2010bq,Heckman:2008rb} that asymptotic freedom of observable matter fields suggests that the Standard Model brane is contractible. From this perspective, it is natural that the axion is localized.  For the remainder of this section we will assume that this is the case.

 We will now argue, based on a generalization of the Weak Gravity Conjecture (WGC), that for the localized axion in the dark dimension scenario, one has
\begin{align}
    f_a\lesssim M_5\sim 10^9-10^{10}{\, \rm GeV} .
    \label{eq:fbound}
\end{align} 

To argue for this bound, we now explain how the WGC might be strengthened in the case that the charged states and gauge fields are localized on a brane. We will then use the axion version of this to arrive at the above bound.


To arrive at the bound \eqref{eq:fbound}, let us consider a 5 dimensional theory where the fifth dimension has an arbitrarily large size $L$. Consider a 4D brane in this setup, localized in the fifth dimension and supporting some $U(1)$ gauge field on its worldvolume. We can apply the usual WGC to the particles localized on this brane. In this setup, the WGC implies that there exists a charged particle with charge $q$ and mass $m$ satisfying
$${m\over M_p}\leq g$$
where $M_p$ is the Planck mass in four dimensions. We can easily write this bound in terms of the 5 dimensional Planck scale using the relation $M_p^2=M_5^3L$:
\begin{align}
{m\over M_5}\leq g \ (M_5 L)^{\frac{1}{2}}.
\label{eq:wgcL}
\end{align}
So far, we have just rewritten the usual WGC in terms of five dimensional quantities. Because $m$ and $g$ are independent of $L$, if there is no potential for $L$ we can consider this for arbitrary value of $L$.  Here we will assume that even if there is a potential for $L$ this bound continues to hold for abitrary $L$, as long as $L$ is not too small compared to $M_5$.    We can therefore consider the most extreme version of the above bound, where we set  $L\sim M_5^{-1}$. Plugging this into \eqref{eq:wgcL}, we arrive at the bound
$${m\over M_5}\leq g.$$
We therefore see that for localized objects one expects the higher dimensional Planck scale to enter the WGC inequalities.  Applying this to the axion we find
$f_a\lesssim M_5\sim 10^{9}-10^{10} \ {\rm  GeV}$
 as in \eqref{eq:fbound}.    Note that the bound $f_a<M_5$  for localized axions is commonly assumed in large extra dimension scenarios and here we have given a motivation for it based on the WGC.

Another way to argue for \eqref{eq:fbound} is as follows.  Suppose we had an axion that propagates in the entirety of the five dimensional bulk. The 5D action contains a kinetic term for this axion:
$$S_5 \supset {1\over 2} \int d^5x\  f_5^3 (\nabla \theta )^2.$$
Using the standard WGC for axions, we have that
\begin{align}
    f_5\lesssim M_5.
    \label{eq:wgc5d}
\end{align}
Now suppose that there is a mechanism for localizing this axion to the SM brane. Then to obtain the four-dimensional axion decay constant, we can integrate the axion kinetic term over the large extra dimension. Assuming the axion field only has support over a region of size $\mathcal{O}(1/M_5)$ in the fifth dimension, we arrive at
\begin{align}
    f_a^2 =\frac{f_5^3}{M_5}.
\end{align}
%
Using \eqref{eq:wgc5d}, we learn that
$f_a\lesssim M_5$ for localized axions. In the dark dimension scenario, we have\footnote{One can also substantiate the bound \eqref{eq:fbound} in string theory: consider, for example, compactifying type IIB on the resolution of $\mathbb{C}^3/\mathbb{Z}_3$. In this case, the radial part of $C_4$ integrated over the blow-up cycle is localized near the origin.} $M_5 \sim 10^9-10^{10}$ GeV.  Using \eqref{eq:fbound}, we can ask what observational constraints exist for the QCD axion. 

The most immediate bound comes from the fact that the QCD axion couples to photons. The axion-photon coupling is given by
\begin{align}
    g_{a\gamma\gamma} = \frac{\alpha}{2\pi f_{a}},
\end{align}
where $\alpha$ is the fine-structure constant of electromagnetism at low energies. As discussed in \S\ref{sec:expbounds}, astrophysical tests and observations put bounds on $g_{a \gamma \gamma}$. For general (not necessarily QCD) axions, these constraints can vary as a function of the axion mass, but for the QCD axion, the most stringent bounds on $g_{a\gamma \gamma}$ comes from the observational requirement that stars do not cool too quickly (see \S\ref{sec:expbounds}). In particular, bounds on neutron stars lead to the limit \cite{Buschmann:2021juv} 
\begin{align}
   f_a \gtrsim    3.5 \times 10^{8}. \ \text{GeV}.
\end{align}
A 4D decay constant in the region specified by \eqref{eq:fbound} gives a theoretical bound on the axion-photon coupling
\begin{align}
    f_a \lesssim 10^{9} - 10^{10} \ \text{GeV}.
    \label{eq:fpred}
\end{align}

This leads us to a striking prediction: if the dark dimension scenario describes our universe, then a QCD axion localized to the Standard Model brane must have a decay constant roughly $f_a \sim 10^9 - 10^{10}$ GeV. Note that this implies that the thickness of the brane is of the order $\ell_5 \sim 10^{-23} \ \text{cm}$.

It is interesting to think about the dark matter composition of such a universe. Coherent oscillations of axions in the early universe contribute to the overall dark matter density. For a general axion-like particle that starts oscillating before matter-radiation equality, the fractional energy density is given by
\begin{align}
    \Omega_a \approx \frac{1}{6} (9\Omega_r)^{3/4} \left(\frac{m_a}{H_0}\right)^{1/2} \bigg\langle \left( \frac{\phi_i}{M_p} \right)^2 \bigg\rangle
\end{align}
where $\Omega_r$ is the fractional energy density due to radiation, $m_a$ is the axion mass, and $\phi_i$ is the initial axion field displacement. For $\mathcal{O}(1)$ values of $\frac{\phi_i}{f_a}$, the energy density scales like $\sqrt{m_a} f_a^2$. One can arrange for a smaller dark matter contribution by tuning $\phi_i$ to be small, or by invoking an anthropic argument to set the initial misalignment angle.

For the QCD axion, the relationship between the decay constant $f_a$ (in GeV) and the mass is fixed, so that the fraction of the total observed dark matter density can be written only in terms of the decay constant and the initial misalignment angle.  With a standard (high) reheat temperature, taking into account the temperature dependence of the QCD axion mass, assuming that $f_a \gtrsim 10^{15} \, \text{GeV}$, and assuming that the QCD axion started oscillating at $T>1 \, \text{GeV}$, the contribution to the total dark matter relic density is \cite{Marsh:2015xka}:
\begin{align}
    \Omega_a h^2 \sim 2 \times 10^4 \left( \frac{f_a}{10^{16} \ \text{GeV}}\right)^{7/6} \langle \theta_{\text{initial}}^2 \rangle 
    \label{eq:QCDdm}
\end{align}
where $\theta_{\text{initial}}$ is the initial misalignment angle (taking values between $0$ and $2\pi$) and $h$ is the dimensionless Hubble parameter. $\Omega_a h^2 = 0.12$ would correspond to the axion making up one hundred percent of the observed dark matter \cite{Planck:2015fie}. 

For the values of decay constants we consider (see \eqref{eq:fpred}), the observed dark matter density doesn't pose much of a constraint: using \eqref{eq:QCDdm}, for a decay constant $f_a \sim 10^9-10^{10} \ \text{GeV}$, the fraction of the overall dark matter density would be about $0.1-1 \%$.

\subsection{Case 2: Axions propagate in the 5d bulk}
We now consider the case in which the axion exists as a bona fide 5-dimensional bulk field. In this case, the simplest string theory compactifications indicate \cite{Svrcek:2006yi} that the axion decay constant will be around the GUT scale, which poses a problem for cosmology. Nevertheless, we will now consider whether the QCD axion in such a scenario is allowed experimentally. 

We consider compactifying this 5D theory on a circle of radius $R$ to obtain an axion in four spacetime dimensions. 

The relationship between the axion decay constant in 5D and the decay constant in 4D is
\begin{align}
    f_{a}^2 = R \, f_{5}^3.
\end{align}

In 5D, we expect the axion to have a decay constant set by the geometry of the internal dimensions, so that (see \S\ref{sec:review})
\begin{align}
    f_{5} \sim \frac{M_5}{r^{2p/3}}
\end{align}
where $M_5$ is the 5-dimensional Planck mass and $r^p$ represents the characteristic volume of the internal cycle giving rise to the axion, measured in string units. 

This in turn means that if the internal geometry has sufficiently small curvature so as to keep the $\alpha'$ expansion under control, then the 4-dimensional decay constant is bounded from above by the 4-dimensional Planck mass:
\begin{align}
    f_a \sim \frac{M_p}{r^{p}}.
    \label{eq:f4d_bulk}
\end{align}
just as in \eqref{eq:fscaling}. Since this axion decay constant is only bounded by the 4D Planck mass, it falls into the same standard parameter space as usual axions from field theory. From this perspective, the bulk QCD axion is not constrainted by axion-photon coupling bounds.

However, in the case that the axion propagates in five dimensions, we must also consider the fact that it comes with a tower of Kaluza-Klein axions with masses set by the radius of the fifth dimension \cite{Dienes:1999gw}:
\begin{align}
    M_{nm}^2 = m_{a}^2 \left( r_n r_m +  \frac{n^2}{m_{a}^2 R^2}   \delta_{nm} \right)
\end{align}
with $m_{a} = \frac{\Lambda_{\text{QCD}}^2}{f_a}$ and $r_n$ the couplings of the axions in the tower to the gauge field. Note that the existence of the Kaluza-Klein modes means that the lightest axion in the tower need not be the QCD axion. If the KK scale is lower than $\frac{\Lambda_{\text{QCD}}^2}{f_a}$, then the lightest axion is the lightest KK mode.

An important point to address is whether the existence of an entire Kaluza-Klein tower of axions is in conflict with any observational bounds. Here, we will consider the constraints on the axion-photon couplings of such a tower. One might expect that even if the axion-photon coupling of the QCD axion on its own is too weak to be within reach of current and future experiments, the cumulative effect of the tower might be observable. Each of the axion-like particles in the tower has an axion photon coupling
\begin{align}
    g_n \approx \frac{\alpha_{\text{EM}}}{2\pi f_a}
\end{align}
and so each of these axions individually is unconstrained by axion-photon coupling bounds if $f_a$ is large enough. 

However, we should ask what the cumulative effect of having a tower of axion-like particles is. The effective coupling to which a given axion-photon experiment is sensitive to is
\begin{align}
    g_{\text{eff}} = \sqrt{ \sum_{n|m_n \leq m_{\text{exp}}} g_n^2}
\end{align}
where $m_{\text{exp}}$ is the mass threshold to which a given experiment is sensitive. 

Suppose for the moment that a powerful enough experiment were sensitive to the photon couplings of axions with arbitrarily high masses, i.e. that $m_{\text{exp}} = M_5$. Then the number of axions that this experiment is sensitive to is $N \approx \frac{M_5}{M_{\text{KK}}}$, where $M_{\text{KK}} \approx 6.6 \, \text{meV}$ for the dark dimension scenario. Since all axions in the KK tower have the same $g_n \equiv g$, the effective coupling becomes

\begin{align}
    g_{\text{eff}}^2 \approx R M_5 \left(\frac{\alpha}{2\pi f_a}\right)^2 =  \left(\frac{\alpha}{2\pi} \right)^2 \frac{M_5}{f_5^3},
\end{align}
so that the relevant scale is the five-dimensional decay constant, rather than the four-dimensional one. An experiment with arbitrarily high-mass sensitivity would therefore see the five-dimensional decay constant, and we would return to the conclusions reached in the previous section.

However, in reality experiments are only sensitive to light axions up to some smaller threshold $m_{\text{exp}} \ll M_5$. In this case, we get
\begin{align}
    g_{\text{eff}}^2 \approx R \, m_{\text{exp}} \left(\frac{\alpha}{2\pi f_a}\right)^2 =  \left(\frac{\alpha}{2\pi} \right)^2 \frac{m_{\text{exp}}}{f_5^3},
\end{align}
which is not any more constraining than the constraints on having a single four-dimensional axion with decay constant $f_a$, because $m_{\text{exp}}/M_{\text{KK}}$ is not parametrically large (for the CAST \cite{CAST:2017uph} experiment, for example, one has $m_{\text{exp}}/M_{\text{KK}} \approx 3$).

We therefore conclude that in the case that the QCD axion is a bulk field in the five-dimensional theory, neither the QCD axion itself nor its KK copies are constrained by observational bounds on axion-photon couplings.

\section{Conclusions}

In this work, we have considered the phenomenological constraints on QCD axions in the dark dimension scenario. We have seen that the case in which the QCD axion is localized on the Standard Model brane (which is natural from the perspective of SM brane phenomenology \cite{Heckman:2008rb,Heckman:2010bq}) solves the old problem that the QCD axion's decay constant is observationally constrained: $10^8-10^9{\, \rm GeV}\leq f_a\leq 10^{11}-10^{12}{\, \rm GeV}$. In the setup of the dark dimension, this scale arises naturally as the Planck mass of the five-dimensional theory: $M_5\sim 10^{10}-10^{11}{\rm GeV}\sim \Lambda^{\frac{1}{12}}$ (where  $\Lambda$ is the dark energy in Planck units). As we argue in \S\ref{sec:axionDD}, this scale bounds the axion decay constant, providing a simple mechanism for why $f_a$ falls in the experimentally allowed range. This mechanism is a realization of the idea put forth in \cite{Svrcek:2006yi} that an axion localized on the SM brane could help to lower the QCD axion decay constant to the phenomenolgically allowed range.

Moreover, the fact that the QCD axion mass $m_a\sim 1-10{\, \rm meV}\sim \Lambda^{1/4}$ coincides with both the mass scale of the neutrinos and the tower mass scale for the dark gravitons in this scenario is somewhat unexpected.  However, parameterically this would always work for arbitrary $\Lambda$, if we use the fact the $\Lambda_{QCD}\sim \Lambda^{2/12}$ and $m_a\sim \Lambda_{QCD}^2/f_a $.
Indeed it is quite satisfactory that in the dark dimension scenario the natural scales in the universe end up being clustered with specific simple powers of the dark energy:
\begin{align*}
    \Lambda^0 &\sim M_p \sim 1\\
    \Lambda^{\frac{1}{12}} &\sim M_5, f_a,\Lambda^{\rm Higgs}_{\rm inst.}\sim 10^{-10} \\
    \Lambda^{\frac{2}{12}} &\sim \Lambda_{\rm QCD}, \alpha \Lambda_{\rm weak} \sim 10^{-20}\\
    \Lambda^{\frac{3}{12}} &\sim m_\nu, m_a, m_{{\rm dark \,  tower}} \sim 10^{-30}\\
    \Lambda^{\frac{6}{12}} &\sim H_0\sim \tau_{\rm now}^{-1}\sim 10^{-60}\\
    \Lambda^{\frac{12}{12}} &= \Lambda \sim 10^{-120}
\end{align*}
As discussed in \cite{Montero:2022prj} the dark tower mass scale is related to the dark energy $m\sim \Lambda^{1/4}$ and the 5D Planck scale follows from the existence of the dark dimension and its size goes as $\Lambda^{1/12}$, which was suggested to be related to the Higgs potential instability above that scale \cite{Montero:2022prj}.
The identification of the Standard Model scale with $\Lambda^{1/6}$ in the above is more tentative:
The Weak scale may be related to demanding lack of hierarchy in the neutrino sector-that the sterile and active neutrinos have similar masses \cite{Montero:2022prj}.   In this regard it is amusing to note the apparent accidental equality of QCD axion mass scale and the neutrino mass scale.
In particular the axion mass and the neutrino mass (which is obtained by seesaw-like mechanism coupling to bulk sterile neutrinos \cite{Dienes:1998sb,Arkani-Hamed:1998wuz}) are both given essentially by the same expressions:
$$m_a\sim \frac{\Lambda_{\rm QCD}^2}{f_a}\sim \frac{(\Lambda^{2\over 12})^2}{\Lambda^{1\over 12}}\sim \frac{(\alpha \Lambda_{\rm weak})^2}{M_5} \sim m_\nu $$

The main result of this work is that in the case that the QCD axion is localized on the SM brane, the QCD axion decay constant $f_a$ is within the range of the next round of axion searches. In the context of the dark dimension scenario, this leads us to anticipate exciting discoveries ahead in upcoming experiments.


\subsubsection*{Acknowledgments} 
 We thank Liam McAllister, Miguel Montero, Jakob Moritz, and Matt Reece for valuable discussions.
This work is supported in part by a grant from the Simons Foundation (602883,CV), the DellaPietra Foundation, and by the NSF grant PHY-2013858.

\bibliographystyle{utphys}
\bibliography{refs}

\end{document}